\documentclass[11pt,a4paper,notoc]{article}

\usepackage{jheppub}
\usepackage{amsmath}         
\usepackage{amssymb}          
\usepackage{bbold}
\usepackage{ulem}
\usepackage{cancel}
\usepackage{afterpage}
\usepackage{youngtab}
\usepackage{multirow}
\usepackage[capitalise]{cleveref}
\usepackage{slashed}
\usepackage{booktabs}
\usepackage{subcaption}
\usepackage{comment}
\usepackage{rotating}
\usepackage{url}
\usepackage{physics}
\usepackage{slashbox}
\usepackage{nicematrix}
\usepackage{tikz}
\usepackage{tikz-feynman}
\usepackage{textpos}
\usepackage{feynmp-auto}
\usepackage{float}
\definecolor{linkcolor}{rgb}{0,0,0.6}
\definecolor{mygreen}{rgb}{0,0.6,0}
\definecolor{ballblue}{rgb}{0.13, 0.67, 0.8}

\crefname{section}{sec.}{secs.}
\crefname{table}{Tab.}{Tabs.}
\crefname{figure}{Fig.}{Figs.}
\crefname{equation}{Eq.}{Eqs.}
\crefname{appendix}{Appendix}{Appendix}

\def\beq{\begin{equation}}
\def\eeq{\end{equation}}

\def\gsim{\raise0.3ex\hbox{$\;>$\kern-0.75em\raise-1.1ex\hbox{$\sim\;$}}}
\def\lsim{\raise0.3ex\hbox{$\;<$\kern-0.75em\raise-1.1ex\hbox{$\sim\;$}}}

\makeatletter
\gdef\@fpheader{} 
\makeatother

\usepackage[utf8]{inputenc}

\title{One loop renormalization of 5D gauge-Yukawa theories}

\author[a,b]{Giacomo~Cacciapaglia,}
\affiliation[a]{Laboratoire de Physique Theorique et Hautes Energies LPTHE, UMR 7589, Sorbonne Université \& CNRS, 4 place Jussieu, 75252 Paris Cedex 05, France}
\affiliation[b]{Quantum Theory Center (QTC) \& D-IAS, Southern Denmark Univ., Campusvej 55, 5230 Odense M, Denmark}
\emailAdd{cacciapa@lpthe.jussieu.fr}
\author[c]{Wanda~Isnard,}
\affiliation[c]{Université Claude Bernard Lyon 1, CNRS/IN2P3, IP2I UMR 5822,  4 rue Enrico Fermi, F-69100 Villeurbanne, France}
\emailAdd{wanda.isnard@ens-lyon.fr}
\author[d]{Roman Pasechnik,}
\emailAdd{roman.pasechnik@fysik.lu.se}
\author[d]{Anca~Preda,}
\emailAdd{anca.preda@fysik.lu.se}
\affiliation[d]{Department of Physics, Lund University, SE-223 62 Lund, Sweden}

\begin{document}

\abstract{The common lore dictates that extra dimensional theories loose predictive power at energies just above the compatification scale, due to the power-law running of bulk coupling. We show that five-dimensional gauge-Yukawa theories can be valid up to arbitrarily high scales, provided:
\begin{enumerate}
    \item A finite number of terms are required to absorb power-law divergences;
    \item All power-law running couplings flow to UV fixed points.
\end{enumerate}
By explicitly computing bulk and localized divergences for a gauge-Yukawa theory on $\mathcal{S}^1/\mathbb{Z}_2$, we prove the one-loop renormalization properties of Lagrangians containing only interactions that would be renormalizable in four dimensions. The existence of UV fixed points imposes further constraints on the content of the model. Our results provide a consistency check for the high-energy behavior of any 5D theory, and provide a discrimination  between UV consistent models and those that can describe only a handful of Kaluza-Klein modes. Hence, we offer the first concrete step towards an all-order proof of `renormalizability' for gauge-Yukawa theories in five dimensions.}

\maketitle

\section{Introduction}
\paragraph{}
Quantum field theories in more than four spacetime dimensions provide a compelling playground for studying extensions of the Standard Model (SM). First employed as arenas for possible gauge-gravity unification \cite{Kaluza:1921tu,Klein:1926tv} and later required for the consistency of superstring theory, it was realized that extra dimensions could be relevant at scales as low as the TeV \cite{Antoniadis:1990ew}. Higher-dimensional field theories, in fact, entail one of the straightforward extensions of space-time symmetries, besides supersymmetric theories. They can address various questions about symmetry breaking dynamics, coupling unification, emergence of mass hierarchies, and the origin of yet-to-be-discovered UV physics. In phenomenologically viable scenarios, the extra dimensions are typically compactified either via boundaries or warped geometry \cite{Randall:1999vf}, so that at sufficiently low energies an effective four-dimensional (4D) theory emerges. This generates a specific new energy scale, where modifications to the 4D behavior emerge, and boundaries when the extra dimensional space geometry is based on orbifolds \cite{Hebecker_2002}. Therefore, the resulting spacetime is characterized by a bulk region, where extra-dimensional fields propagate, and boundaries, where lower-dimensional localized interactions and additional degrees of freedom may live as long as they respect the locally preserved symmetries. 
These features open the question of the validity of the extra dimensional theories and the sensitivity of the computable results to unknown UV corrections. The problem is twofold. On the one hand, couplings in the bulk (including gauge ones) are affected by power-law corrections \cite{Contino:2001si,Hebecker:2002vm} and may become non-perturbative shortly above the compactification scale, hence deeming the extra dimensional model a mere effective description of a more UV complete theory (such as string theory). On the other hand, higher dimensional operators may be required to regularize divergences, both in the bulk and on the boundaries \cite{Contino:2001si,Hebecker_2004,Ghilencea:2006qm}.
In this paper, we critically analyze the conditions under which an extra dimensional theory can be considered `renormalizable' \cite{Gies_2003,Morris_2005}. This consists in the absence of higher dimensional operators in the bulk required by loop divergences, and the taming of the power-law running of the couplings. The latter is achieved by requiring the presence of UV fixed points for such couplings. We focus on 5D theories, where both conditions are most likely to occur, while theories with more dimensions have been proven to be always non-renormalizable (see, e.g., \cite{Ghilencea:2006qm}).
This issue is of particular interest in Grand unification theories (GUTs) in 5D, where the power-law running has been used to obtain accelerated quantitative unification \cite{Dienes_2003} and asymptotic unification via fixed points \cite{Cacciapaglia21}.

Upon reduction to a 4D effective field theory, the higher-dimensional fields decompose into Kaluza–Klein (KK) towers \cite{Perez_Lorenzana_2005} with masses and interactions determined by both the geometry and the boundary conditions. As a result, the particle content relevant at any given energy scale changes in a highly constrained way. This can substantially modify the dependence of gauge, Yukawa, and scalar couplings on energy, compared to the familiar 4D case \cite{Dienes_2003}.
It is well known that in 4D theories, the evolution of couplings, governed by the renormalization group equations (RGEs), has a logarithmic dependence on the renormalization scale. However, this behavior changes to a power-law dependence when theories are formulated in higher dimensions, precisely due to the presence of the KK modes that accelerate the running. This is particularly relevant in the context of Grand unification \cite{Cacciapaglia21}. Traditional 4D unification scenarios are built around the logarithmic running of couplings and threshold effects associated with heavy states. In extra dimensional models, by contrast, an increasing number of KK levels contribute, and the high-energy behaviour is no longer generically described by a single unification point \cite{Bajc_2016}. This motivates revisiting the idea of unification from the viewpoint of higher-dimensional gauge theories, and in particular exploring under which conditions it can be realized.

In this work we focus on 5D gauge theories compactified on a $\mathcal{S}^1/\mathbb{Z}_2$ orbifold, and we aim at determining the minimal Lagrangian and field content leading to a renormalizable theory, as defined above. We focus on boundary conditions consistent with an orbifold as it is the only case that preserves the 5D gauge symmetry at high scales, see e.g. \cite{Cacciapaglia:2023kyz}. We limit our analysis at one loop, hence providing the first concrete step towards an all-order proof of renormalizability. Henceforth, we study the structure of the required bulk and localized counterterms,  as well as the running of bulk gauge, Yukawa, and scalar quartic couplings. The presence of UV fixed points for the power-law running couplings imposes additional requirements on the field content and symmetries of the bulk theory and on the localized Yukawa couplings. We find that a theory containing 4D-renormalizable interactions in the bulk and boundaries may fulfill the renormalizability criteria in 5D. The effect of higher-loop orders is highly non-trivial, and it may require the inclusion of additional higher dimensional operators. However, a detailed analysis is required, which we leave for future investigation: the main point being that the presence of logarithmic running could be acceptable, as long as the 5D theory remains valid until scales much higher than the compactification one. The impact of higher-loop orders on the fixed points is less important as long as the fixed point values are perturbative. Also, fixed points in 5D have received some support from non-perturbative analyses \cite{Gies_2003,Codello:2016muj,DeCesare:2021pfb,Pastor-Gutierrez:2024huj}.

The manuscript is organized as follows. In Section \ref{sec:bulkdiv}, we analyze the structure of one loop bulk divergences in 5D gauge-Yukawa theories. Using power-counting arguments, we show that at one-loop the divergent operator structure in 5D coincides with the one of 4D theories, with divergences enhanced by the power-law sensitivity to the UV cutoff. We also discuss how this conclusion could be extended to higher loops. Section \ref{sec:locdiv} is devoted to the study of divergences localized on the orbifold fixed points. We begin with a detailed study of a pure Yang-Mills theory compactified on a $\mathcal{S}_1 / \mathbb{Z}_2$ orbifold, illustrating how the boundary terms arise from the breaking of translational invariance along the fifth dimension. We then extend the analysis to include fermions, scalars and Yukawa interactions, computing all the localized divergent contributions at one loop. We show that these divergences can be consistently absorbed into a finite set of renormalizable 4D operators localized on the boundary. The results are summarized in Section \ref{sec:summary}, where we present the complete form of the one loop renormalized boundary Lagrangian $\mathcal{L}_4$. In Section \ref{sec:bulkRGEs} we compute the RGEs of bulk couplings and investigate the general conditions under which 5D theories admit UV fixed points.  Section \ref{sec:RGEloc} extends the RGE analysis to localized Yukawa interactions, as other terms feature typical 4D logarithmic running. We conclude in Section~\ref{sec:concl}. Throughout this work, we use the metric convention $(+,-,-,-,-)$.

\section{One loop renormalization in 5D} \label{sec:bulkdiv}

We start from the most general $d$-dimensional Lagrangian containing spin $0$, $1/2$ and $1$ fields, with the lowest dimensional interactions (leading to a theory renormalizable in $d=4$):
\begin{equation} \label{eq:LagrGen}
\begin{aligned}
    \mathcal{L}_d= & -\frac{1}{4}\ F^{a,MN}F^a_{MN}-\frac{1}{2 \xi} \left( \partial_M A^{a,M} \right)^2+ i \overline{\Psi} \Gamma^M \mathcal{D}_M \Psi + (\mathcal{D}_M \Phi^\dagger) (\mathcal{D}^M \Phi) \\
    & - y_d\ \overline{\Psi}_1 \Psi_2 \Phi - \lambda_d\ (\Phi^\dagger \Phi)^2\,,
\end{aligned}
\end{equation}
where the indices $M(N)=\{\mu,5 \dots d\}$ and $F^a_{MN}=\partial_M A^a_N-\partial_N A^a_M+ g_d\ f^{abc}A^b_M A^c_N$, $\mathcal{D}_M = \partial_M - ig_d\ A_M^a T^a$. The gauge-fixing term corresponds to a 5D generalized Lorentz gauge condition.
As the Lagrangian has mass dimension $d$, the fields have canonical mass dimensions
\begin{equation}
    [A_M] = m^{(d-2)/2}\,, \quad [\Psi] = m^{(d-1)/2}\,, \quad [\Phi] = m^{(d-2)/2}\,, 
\end{equation}
yielding the following mass dimensions for the couplings:
\begin{equation}
    [g_d] = [y_d] = m^{(4-d)/2}\,, \quad [\lambda_d] = m^{(4-d)}\,.
\end{equation}
As expected, for $d=4$ the three couplings are dimensionless, leading to a renormalizable theory \cite{tHooft:1971akt,tHooft:1972tcz,Piguet:1995er}. Instead, for $d>4$ they have negative mass dimension, which can be naively interpreted as a sign of non-renormalizability. We recall that for a compact extra dimensional space, one can define effective dimensionless couplings as follows:
\begin{equation}
    g = \frac{g_d}{\sqrt{\mathcal{V}}}\,, \quad y = \frac{y_d}{\sqrt{\mathcal{V}}}\,, \quad \lambda = \frac{\lambda_d}{\mathcal{V}}\,, 
    \label{eq:effectivecouplings}
\end{equation}
where $\mathcal{V}$ is the volume of the $d-4$ dimensional compact space.

\subsection{Bulk divergences at one-loop}

To probe the renormalizability of the Lagrangian \eqref{eq:LagrGen}, we first inspect the structure of the divergences at one loop. As they originate from very short distances in the bulk, we can consider all dimensions to be infinite, and work in full $d$-momentum space, where propagators have the familiar structure
\begin{equation}
    G_x(p^M) = \frac{i\  n_x(p^M)}{p^2-M_x^2}\,,
\end{equation}
where $p^2 = p^M p_M$ and the numerator depends on the spin of the field.

A generic loop in $d$ dimensions will contain an integral of the form
\begin{equation}
    I_d \sim \int \frac{\dd^d k}{(2\pi)^d}\frac{f(k^M,k^2)}{\Delta (k^2)}\,.
\end{equation}
After performing the Wick rotation to go to Euclidean space, one obtains
\begin{equation}
    I_d \sim \frac{\Omega_d}{(2\pi)^d} \int k^{d-1} \dd k\ \frac{f}{\Delta(k^2)},
\end{equation}
where, in the limit of large $k$, 
\begin{equation}
    \Delta(k^2) \sim k^{2n}\,, \quad f(k) \sim k^{2m}\,.
\end{equation}
In the above limits, $n$ counts the number of propagators in the diagram, while $2m$ is the largest power of $k$ in the numerator. Note that the numerator will always be an even power of $k$.
In $d=4$, the integral is finite as long as $n-m>2$, while a logarithmic divergence emerges for $n-m=2$. For $d=5$, $n-m>2$ still leads to finite integrals, while $n-m=2$ gives a linear divergence. This simple argument proves, in fact, that $d=4$ and $d=5$ share the same divergent structure: a 4D theory that is renormalizable, therefore, will also be renormalizable in 5D at one loop, in the sense that all the (linear) divergences can be reabsorbed by the existing couplings.
This is the case for the Lagrangian \eqref{eq:LagrGen}.
This is not the case for $d\geq 6$: the $n-m=3$ case will also diverge. One explicit case is given by box diagrams that generate four-fermion interactions, which have $n=4$ and $m=1$: they remain finite in $d=4,5$, give log divergences in $d=6$ and power-law divergences in larger dimensions.

\begin{table}[htb]
\centering
\begin{tabular}{l||c|c|c|c|l|}
$n-m$ & \multicolumn{4}{c|}{$d =$} & Examples \\
 & $4$ & $5$ & $6$ & $7$ & \\
 \hline\hline
$1$ & $\Lambda^2$ & $\Lambda^3$ & $\Lambda^4$ & $\Lambda^5$ & Scalar masses \\ \hline
$2$ & log & $\Lambda$ & $\Lambda^2$ & $\Lambda^3$ & Coupling renormalization \\\hline
$3$ & finite & finite & log & $\Lambda$ & Four-fermions, $(\Phi^\dagger \Phi)^3$ \\\hline
$\geq 4$ & finite & finite & finite & finite & \\\hline
\end{tabular}
\caption{Divergence structure of loops in various dimensions as a function of $n-m$.} \label{tab:bulk}
\end{table}

A summary of the divergence structure for $d=4,5,6,7$ is shown in Table~\ref{tab:bulk}. The pattern that emerges is that $d=2p$ and $2p+1$ for integer $p$ share the same divergent integrals. It is also straightforward to check that the Lorentz structure of each divergent loop is the same in the two cases.

This simple analysis allows us to prove that the Lagrangian in \eqref{eq:LagrGen} can be renormalized in both $d=4$ and $d=5$ at one loop. This property is lost for $d\geq 6$, where additional counterterms in the bulk are necessary.

\subsection{Extension to higher loops}

The power counting argument can be straightforwardly extended to higher loops~\cite{Zimmermann:1968mu,Lowenstein:1975rg}. A typical $N$-loop integral in the limit of large loop momenta $k_i$ can be written as:
\begin{equation}
    I_N \sim \left( \Pi_{i=1}^N \int {\dd}^d k_i \right)\ \frac{f(k_i^2, k_j\cdot k_l)}{\left(\Pi_{i=1}^N (k_i^2)^{n_i}\right) \left(\Pi_{l>j} ((k_l-k_j)^2)^{n_{lj}} \right)}\,,
\end{equation}
where $n_i$ and $n_{lj}$ are integers, while the function at the numerator depends on the details of the vertices. This general master integral contains two types of divergences:
\begin{itemize}
    \item Sub-divergences due to the superficial divergence of a sub-set of momentum integrals. They can be renormalized by counterterms at loop order $< N$.
    \item Global divergence at $N$-loops, which can be measured by scaling all $k_i \to \lambda\ k_i$.
\end{itemize}
The latter is responsible for the appearance of genuine $N$-loop divergences, which depend on the scaling of the integrand. The latter is always an even power, $\lambda^{2n_I}$.
Hence
\begin{itemize}
    \item[{\it i})] For $2n_I = 4 N$, the integral has a superficial log-divergence in $d=4$.
    \item[{\it ii})] In $d=5$, the same integrals feature a power-law divergence $\Lambda^{N}$.
    \item[{\it iii})] Integrals with $4N < 2n_I < 5N$ are finite in $d=4$, however they have power-law divergences in $d=5$.
    \item[{\it iv})] For even $N$, $2n_I = 5N$ integrals, finite in $d=4$, generate new log-divergences in $d=5$.
\end{itemize}
Taken at face value, observations {\it iii} and {\it iv} would imply that new counterterms are needed for $d=5$ starting with log-divergent ones at two loops and power-law divergent ones at three loops. However, this naive conclusion requires a more detailed analysis to be confirmed, for instance along the lines of Zimmermann's forest equations \cite{Bogoliubov:1957gp,Zimmermann:1969jj} to correctly identify the divergences and the role they play in the renormalization of the theory. We leave this analysis for future work. 

This caveat certainly holds for power-law divergences, which are characteristic of $d=5$. However, for even number of loops, log-divergences also appear, which may lead to genuinely new operators in the bulk.  As they run logarithmically, their presence would still allow the 5D theory to be valid up to scales much higher than the compactification scale, hence they would be tolerable.

\section{Localized divergences and operators} \label{sec:locdiv}

Any realistic model in flat 5D requires the compactification of the extra dimension on a finite interval. One of the main reasons lies in the necessity of obtaining chiral 4D zero modes out of the non-chiral 5D fermion fields. The compact space, therefore, must have boundaries, on which one can localize 4D interactions. Localized divergences also arises at loop level, hence requiring the presence of boundary counterterms. Henceforth, the renormalization of any realistic 5D theory must include localized interactions. 

Without loss of generality, we can consider the minimal orbifold $\mathcal{S}^1/\mathbb{Z}_2$, consisting of a circle modded by a $\mathbb{Z}_2$ parity acting on the fifth coordinate $x^5$.~\footnote{As we will see, the localized divergences only depend on the gauge symmetry unbroken on the boundary. Hence, the most general $\mathcal{S}^1/\mathbb{Z}_2\times\mathbb{Z}_2'$ orbifold can be easily obtained by straightforward extension of our results.} In this case, only one boundary exists, as the two borders of the interval are identified by the parity.
The most general Lagrangian, therefore, will contain two pieces: a 5D term $\mathcal{L}_5$ identical to Eq.~\eqref{eq:LagrGen}, and $\mathcal{L}_4$ localized on the 4D boundary. The action, therefore, can be written as~\footnote{For the most general case $\mathcal{S}^1/\mathbb{Z}_2\times\mathbb{Z}_2'$, the action would read
$$
    S_5 = \int {\dd}^5x\ \Big\{ \mathcal{L}_5 + \mathcal{L}_4\  \delta(x^5) + \mathcal{L}_4'\  \delta(x^5-L/2)\Big\}\,,
$$ where $\mathcal{L}_4$ can be obtained by studying $\mathcal{S}^1/\mathbb{Z}_2$ and $\mathcal{L}_4'$ by studying $\mathcal{S}^1/\mathbb{Z}_2'$.}
\begin{equation} \label{eq:S5}
    S_5 = \int {\dd}^5x\ \left\{ \mathcal{L}_5 + \left[ \mathcal{L}_4 \frac{\delta(x^5) + \delta(x^5-L)}{2}\right]\right\}
\end{equation}
defined on an interval $[0,L]$. As $\mathcal{L}_4$ is a 4D Lagrangian in terms of bulk fields evaluated at the boundary, i.e. 4D projections, naive expectation is that the theory may remain renormalizable as long as $\mathcal{L}_4$ only contains interactions that are 4D-renormalizable. We will check this in this section via explicit computation of the localized divergences.

\subsection{One loop renormalization: a pure Yang-Mills case study}

To illustrate the computation, we provide some details in this section about a pure Yang-Mills theory formulated in 5D and compactified on a $\mathcal{S}_1 / \mathbb{Z}_2$ orbifold. 
The bulk Lagrangian is simply given by:
\begin{equation}
    \mathcal{L}_5=-\frac{1}{4}F^{a,MN}F^a_{MN}-\frac{1}{2 \xi} \left( \partial_M A^{a,M} \right)^2\,,
\end{equation}
where the indices $M(N)=\{\mu,5\}$ and $F^a_{MN}=\partial_M A^a_N-\partial_N A^a_M+g_5 f^{abc}A^b_M A^c_N$. The gauge-fixing term corresponds to a 5D generalized Lorentz gauge condition, which was chosen in order to simplify the loop calculations. 

The action of the orbifold projection on the gauge fields can be expressed as
\begin{equation}
    T^a A_\mu^a (x^\mu, -x^5) = P\cdot T^a\cdot P\ A^a_\mu (x^\mu,x^5)\,, \quad T^a A_5^a (x^\mu, -x^5) = - P\cdot T^a\cdot P\ A^a_5 (x^\mu,x^5)\,,
\end{equation}
where $P$ is a transformation generated by the bulk gauge symmetry such as $P\cdot P=1$.
Under a non-trivial $P$, therefore, some generators $T^a$ are odd, hence the bulk gauge group $\mathcal{G}$ is broken to a subgroup $\mathcal{H}$. 

\begin{figure}[tbh]
\centering
\includegraphics[width=0.4\textwidth]{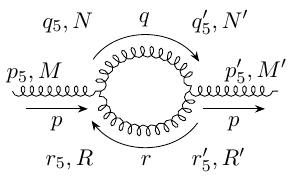}
\caption{One loop diagram contributing to the gauge boson self-energy.}
    \label{fig:gauge_loop}
\end{figure}

The gauge-field propagators are also modified from the usual 4D case due to the orbifold constraints on the fifth dimension. 
In the Feynman gauge ($\xi=1$), for the gauge boson corresponding to the generator $T^A$, these are \cite{von_Gersdorff_2002,Georgi_2001}:
\begin{subequations}
\begin{eqnarray}
    G^A_{\mu \nu}(p,p_5)&=&\frac{-i g_{\mu \nu}}{2(p^2-p_5^2)} (\delta_{p_5,p_5'}+ \eta^A \delta_{-p_5,p_5'})\,, \\
    G_{55}^A(p,p_5)&=&\frac{i}{2(p^2-p_5^2)} (\delta_{p_5,p_5'}-\eta^A \delta_{-p_5,p_5'})\,,
\end{eqnarray}
\end{subequations}
corresponding to the 4D vector $A_\mu$ and the gauge scalar $A_5$, respectively. In the above notation, $p^2 = p^\mu p_\mu$, while $p_5$ denotes the fifth component of the 5D momentum. We also introduced $\eta^A=\pm 1$ corresponding to the eigenvalue of the action of $P$ on the generators: $P \cdot T^A \cdot P = \eta^A T^A$. The 5D propagator can be written in a more compact form as follows
\begin{equation}
    G^A_{MN}(p,p_5)=\frac{-i g_{MN}}{2(p^2-p_5^2)} (\delta_{p_5,p_5'}+ \alpha^{AM}\delta_{-p_5,p_5'}),
\end{equation}
where $\alpha^{A\mu}=\eta^A$ and $\alpha^{A5}=-\eta^A$.
Due to the new structure of the propagators, each one-loop $n$-point function will have two types of contributions: one that conserves the fifth dimensional momentum $p_5$ and one that does not, as translation invariance along $x^5$ is broken by the interval boundaries. The former is associated with bulk divergences to be renormalized by bulk coupling, while the latter corresponds to  localized divergences on the boundaries. To see that explicitly, we can look at the structure of the gauge boson loop in Figure~\ref{fig:gauge_loop} as an example, leading to the following two-point function:
\begin{multline}
    \Pi _{MM'}\sim \delta^{AB} \int \frac{{\rm d}^4q}{(2 \pi)^5} \sum_{q_5} \delta_{p_5-q_5+r_5}\delta_{q'_5-r'_5-p'_5} \ N_{MM'}\;\; \times  \\
     \frac{\delta_{q_5-q'_5}+\alpha^{AN}\delta_{q_5+q'_5}}{q^2-q_5^2}g_{NN'}\frac{\delta_{r_5-r'_5}+\alpha^{BR}\delta_{r_5+r'_5}}{r^2-r_5^2}g_{RR'}\,,
\label{eq:gauge boson loop}
\end{multline}
where $N_{MM'}$ takes into account the vertices of the diagram. Note that the integration over the fifth component of the loop momentum is expressed as a sum due to the compactification,  and the delta functions entail the conservation of the discrete momenta along the extra dimension. For simplicity we omit details of the group structure as well as constant factors.

Multiplying the delta-functions results in four different terms, which schematically take the form:
\begin{subequations}
\begin{eqnarray}
    &\delta_{p_5-p'_5}\,\delta_{NN'}\delta_{RR'}\,,&
    \label{eq:eq1} \\
    & \alpha^N\alpha^R\,\delta_{p_5+p'_5}\,\delta_{NN'}\delta_{RR'}\,,& 
     \label{eq:eq2}\\
    &\alpha^R\,\delta_{p_5+p'_5-2q_5}\,\delta_{NN'}\delta_{RR'}\,,&
     \label{eq:eq3}\\
    &\alpha^N\,\delta_{p_5-p'_5-2q_5}\,\delta_{NN'}\delta_{RR'}\,.&
     \label{eq:eq4}
\end{eqnarray}
\end{subequations}
It is the first two terms, Eqs.~\eqref{eq:eq1} and \eqref{eq:eq2}, that conserve 5D momentum and lead to divergences in the bulk, while Eqs.~\eqref{eq:eq3} and \eqref{eq:eq4} give rise to localized contributions. This is a feature of all loop diagrams and, as a result, the discussion on the renormalizability of the theory needs to be split into two parts: one that investigates the bulk divergences and the corresponding counterterms which need to be added, and one which computes the divergences arising from the localized boundaries. 

Having discussed the bulk divergences in the previous section, in the following we focus on the localized divergences needed to extract the necessary terms in $\mathcal{L}_4$. The integrals on the 4D momenta are done in general $\xi$ gauge and in dimensional regularization at a scale $\mu$. As we will see, only 4D logarithmic divergences emerge, signaling that $\mathcal{L}_4$ does indeed behave like a 4D Lagrangian.

\subsection{Localized renormalization of gauge bosons}
\paragraph{}
We first focus on the renormalization of the gauge bosons, starting from their self-energies -- i.e. corrections to two-point functions. The computation of the bulk and localized divergences have already been carried out in \cite{Cheng_2002,von_Gersdorff_2002}. Here we recomputed the loops, focusing on the localized divergences. 

\begin{figure}[tbh]
\centering
\includegraphics[width=0.9\textwidth]{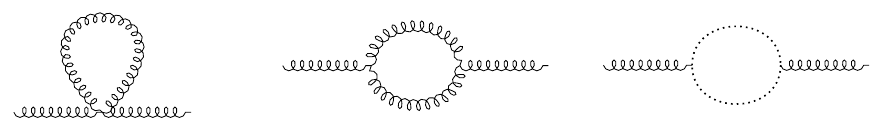}
    \caption{One loop diagrams contributing to the gauge bosons self energy in a pure Yang-Mills theory.}
    \label{selfenergygaugediagram}
\end{figure}

Starting from the two-point function of the 4D vector part $\Pi_{\mu\nu}^\text{loc}$, we see that it receives contributions from the gauge interactions via the three diagrams in  Figure~\ref{selfenergygaugediagram}. 
Adding all the contributions together yields:
\begin{equation}
    \Pi_{\mu \nu}^{\rm loc, gauge}= \frac{g_5^2}{64 \pi^2} \ln \left( \frac{\Lambda^2}{\mu^2} \right) \delta^{AA'}  \left[ (g_{\mu \nu}p^2-p_{\mu} p_{\nu}) \ a_1 + g_{\mu \nu} \frac{p_5^2+p_5'^2}{2} \ a_2 \right]\,,
    \label{eq:selfenergycounterterm}
\end{equation}
where 
\begin{equation}
    a_1 = C \left(\frac{11}{3}-(\xi-1)\right)\,, \qquad a_2 = C\left(4+(\xi-1)\right)\,.
\end{equation}
The normalization $C$ is given by
\begin{equation}
    C= \frac{\eta^A +1}{2} (2 C_2(\mathcal{H}) - C_2(\mathcal{G}))\,.
\end{equation}
Here, $C_2(\mathcal{G})$ is the Casimir of the bulk group, while $C_2(\mathcal{H})$ is the Casimir of the unbroken subgroup the generator $A$ belongs to. In fact, we already see that only generators in $\mathcal{H}$ receive these corrections, as they have $\eta^A =1$. The first term in Eq.~\eqref{eq:selfenergycounterterm} corresponds to a counterterm $\sim F^{A,\mu\nu} F^A_{\mu\nu}$ and for $\xi=1$ its coefficient matches the standard result for a 4D gauge theory. Instead, the second term corresponds to a counterterm $\sim (\partial_5^2 A_\mu^A) A^{A,\mu} + A^A_\mu (\partial_5^2 A^{A,\mu})$. While both are compatible with the orbifold parities, the latter is not gauge invariant. It is well known that a gauge-invariant counterterm can be reconstructed in the ``{\it magic}'' gauge $\xi=-3$, as already observed in 4D \cite{ItzinsonZuber} and in extra dimensional theories \cite{Cacciapaglia:2011hx}. In this gauge, the second term vanishes, $a_2=0$, and we are left with 
\begin{equation}
    \Pi_{\mu \nu}^{\text{loc}, \xi=-3 }= \frac{g_5^2}{64 \pi^2} \ln \left( \frac{\Lambda^2}{\mu^2} \right) \delta^{AA'}\ C\ \frac{23}{3}  (g_{\mu \nu}p^2-p_{\mu} p_{\nu})\,.
    \label{eq:selfenergycounterterm2}
\end{equation}
The two-point function above also receives contributions from bulk scalars but not fermions, giving \cite{Cheng_2002}:
\begin{equation}
    \delta a_1^{\rm scalars} = - \frac{1}{3} \left( \sum_{\rm even} - \sum_{\rm odd} \right) T(r_s)\,, \quad \delta a_2^{\rm scalars} = 0\,,
\end{equation}
where $T(r_s)$ is the index of the representation $r_s$ of the scalar components under the subgroup of $\mathcal{H}$ the generator $T^A$ belongs to.
All in all, we find, for the gauge bosons belonging to the unbroken group $\mathcal{H}$:
\begin{multline}
    \Pi_{\mu \nu}^{\text{loc},\xi=-3} = \frac{g_5^2}{64 \pi^2} \ln \left( \frac{\Lambda^2}{\mu^2} \right)\ \delta^{AA'} (g_{\mu \nu}p^2-p_{\mu} p_{\nu})\ \times \\ \left( \frac{23}{3} \left(2C(\mathcal{H})-C(\mathcal{G})\right) - \frac{1}{3} \left( \sum_{\rm even} - \sum_{\rm odd} \right) T(r_s)\right) \,.
    \label{2pointfunction}
\end{multline}
We also explicitly checked that, in agreement with \cite{von_Gersdorff_2002,Cheng_2002}, no other two-point functions receive localized divergent contributions.

\begin{figure}[tbh]
\centering
\includegraphics[width=0.8\textwidth]{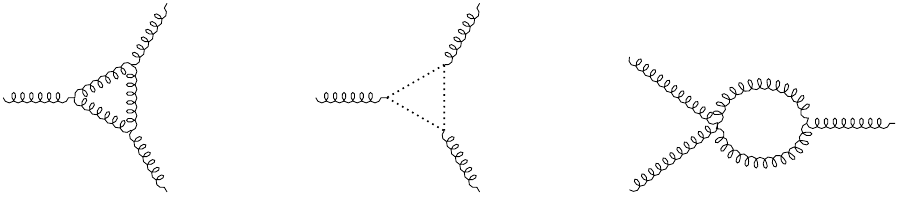}
    \caption{One loop diagrams contributing to the three gauge boson vertex in a pure Yang Mills theory.}
    \label{fig:3pointfunctions}
\end{figure}

Note that the structure of the two-point function $\Pi_{\mu\nu}^{\text{loc}, \xi=-3}$ matches that of a localized gauge-invariant operator $F^A_{\mu\nu} F^{A,\mu\nu}$. Three- and four-point functions should also match, in agreement with localized Ward identities, which should be respected as the correction only emerges for the localized fields corresponding to the unbroken subgroup $\mathcal{H}$.  No other gauge invariant 4D-renormalizable counterterm can be written down in $\mathcal{L}_4$. Nevertheless, to check the consistency of the result, we explicitly computed the localized divergences stemming from the three-point functions represented in Figure \ref{fig:3pointfunctions}. We checked that the localized divergent part of those diagrams is coming only from the $4$D vector bosons $\Pi_{\mu \nu \rho}^{\rm loc}$. The final result reads:
\begin{equation}
    \Pi_{\mu \nu \rho}^{\rm loc}=\frac{g_5^3\  f^{ABC}}{64 \pi^2} \ln \left( \frac{\Lambda^2}{\mu^2} \right) \mathcal{O}_{\mu \nu \rho}\ C \left( \frac{14}{3} + \frac{3}{4}(1-\xi)  \right)\,,
\end{equation}
where $\mathcal{O}_{\mu \nu \rho}=g_{\mu \nu}(p+q)_{\rho}+g_{\nu \rho}(p-2q)_{\mu}+g_{\rho \mu}(q-2p)_{\nu}$ is the usual tensor from tree level trilinear coupling, in the convention where $p$ is the incoming momentum and $q$ is outgoing. Once we include the scalar contributions and consider the ``{\it magic}'' gauge $\xi=-3$ we get:
\begin{equation}
     \Pi_{\mu \nu \rho}^{\text{loc}, \xi=-3}=\frac{g_5^3\ f^{ABC}}{64 \pi^2}\ln \left( \frac{\Lambda^2}{\mu^2} \right) \mathcal{O}_{\mu \nu \rho} \left[  \frac{23}{3} \left(2C(\mathcal{H})-C(\mathcal{G})\right) - \frac{1}{3} \left( \sum_{\rm even} - \sum_{\rm odd} \right) T(r_s) \right]\,,
\end{equation}
in agreement with the coefficient of the two point function in \eqref{2pointfunction}, such that localized Ward identities are respected. Similarly to the previous case, we explicitly checked that all other three-point functions, namely $\Pi_{\mu \nu 5}^{\rm loc}$, $\Pi_{\mu 5 5}^{\rm loc}$ and $\Pi_{5 5 5}^{\rm loc}$, receive no divergent contributions.

The situation is very similar for the four-point function, since it is only $\Pi_{\mu \nu \rho\sigma}^{\rm loc}$ contributing on the boundary, while all other four-point functions, such as $\Pi_{\mu\nu 5 5}$ and $\Pi_{5555}$, vanish.
In particular, a quartic coupling for the gauge-scalar fields associated to the broken generators, which would be allowed by parities, is not generated.

\subsection{Fermions and Yukawa couplings}

\begin{figure}[tbh]
\centering
\includegraphics[width=0.8\textwidth]{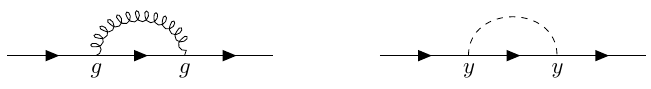}
    \caption{One loop diagrams contributing to the fermion self energy.}
    \label{fig:fermion loops}
\end{figure}

Once fermions are included in  the theory, one needs to consider localized divergences for the fermion couplings, including eventual Yukawa couplings.
We start with the self energy, which has been computed in \cite{Cheng_2002}, by evaluating the diagrams in Figure~\ref{fig:fermion loops}. We identify a logarithmically divergent part, expressed as:
\begin{eqnarray}
    (\Sigma_f)^{\pm}=\frac{1}{64\pi^2}\ln \left( \frac{\Lambda^2}{\mu^2} \right)\left[\cancel{p}\frac{1\pm\gamma_5}{2}b_1-\left(ip_5\frac{1\pm\gamma_5}{2}-ip^{'}_5\frac{1\mp\gamma_5}{2}\right)b_2\right]\,,
\end{eqnarray}
where the upper signs apply to the fermion components with right-handed parity-even, and the bottom ones to left-handed parity-even. The two coefficients read in $\xi$-gauge:
\begin{subequations}
\begin{eqnarray}
    b_1&=&\left(-1-2(\xi-1)\right)\ g_5^2 C(r)- \left( \sum_{\rm even} - \sum_{\rm odd} \right) y_5^2\,, \\
    b_2&=&\left(5+(\xi-1)\right)\ g_5^2 C(r) - \left( \sum_{\rm even} - \sum_{\rm odd} \right) y_5^2\,.
\end{eqnarray}
\end{subequations}
We used the same notation as in \cite{Cheng_2002} where $C(r)$ is the Casimir for fermions in the representation $r$ of the unbroken group $\mathcal{H}$ and, in the Yukawa contribution, the sums cover scalars of different parities. 

\begin{figure}[tbh]
\centering
\includegraphics[width=0.7\textwidth]{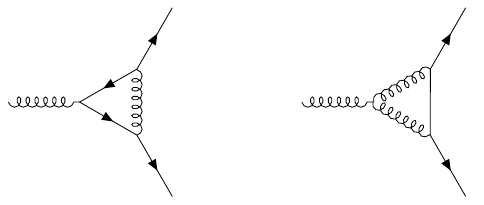}
    \caption{One loop diagrams contributing to the fermion-gauge boson vertex correction.}
    \label{fig:fermion-gauge loops}
\end{figure}

Localized divergences also emerge from corrections to the vertex with gauge bosons, shown in Figure~\ref{fig:fermion-gauge loops}. The vertex correction can be written as:
\begin{eqnarray}
    (\Sigma'_f)^\pm=\frac{1}{64\pi^2}\ln \left( \frac{\Lambda^2}{\mu^2} \right)\ i g_5 \gamma^\mu T^A \frac{1\pm \gamma_5}{2}\ b'_1\,, 
\end{eqnarray}
with
\begin{equation}
    b'_1 = C(r)\left(1+2(\xi-1)\right)g_5^2+C(\mathcal{H})\left(2+\frac{1}{2}(\xi-1)\right)g_5^2 + \left( \sum_{\rm even} - \sum_{\rm odd} \right) y_5^2\,.
\end{equation}
The signs correspond to the chirality of the parity-even component, and the only non-vanishing corrections involve the parity-even generators $T^A$ in $\mathcal{H}$.
We see that in the ``{\it magic}'' gauge
\begin{subequations}
\begin{eqnarray}
    b_1 = - b'_1 &=& 7\ g_5^2 C(r)- \left( \sum_{\rm even} - \sum_{\rm odd} \right) y_5^2\,, \\
    b_2&=& g_5^2 C(r) - \left( \sum_{\rm even} - \sum_{\rm odd} \right) y_5^2\,.
\end{eqnarray}
\end{subequations}
Hence, the counterterms can be expressed in terms of gauge-invariant operators under $\mathcal{H}$, in the form
\begin{equation}
     b_1\ i\bar{\psi} \cancel{\mathcal{D}} \frac{1\pm\gamma_5}{2} \psi + b_2 \left( (\partial_5 \bar{\psi}) \frac{1\pm\gamma_5}{2} \psi - \bar{\psi} \frac{1\mp\gamma_5}{2} (\partial_5 \psi)\right)\,.
\end{equation}

We also computed corrections to the Yukawa couplings, and confirmed the result in \cite{Georgi_2001} that localized divergences vanish, hence no Yukawa counterterms are needed.

\subsection{Scalars and quartic couplings}

In the scalar sector, logarithmic divergences emerge for the self-energies as well as for quartic couplings (setting aside the notorious quadratic divergences of mass terms).

\begin{figure}[tbh]
\centering
\includegraphics[width=0.7\textwidth]{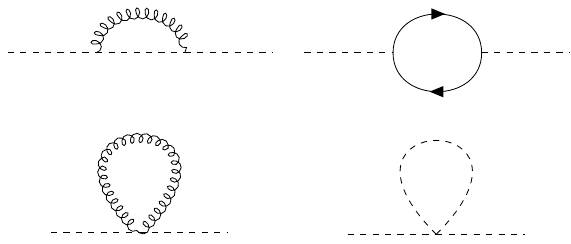}
    \caption{One loop diagrams contributing to the scalar two point function.}
    \label{fig:scalar-loops}
\end{figure}

For the self energy, the divergent contribution of the diagrams in Figure~\ref{fig:scalar-loops} has been computed for an even parity scalar \cite{Cheng_2002}, leading to the general form
\begin{equation}
    (\Sigma_s)^+=\frac{1}{64\pi^2}\ln \left( \frac{\Lambda^2}{\mu^2} \right)\left[p^2\  c_1+\frac{p^2_5+p'^2_5}{2}\ c_2\right]\,,
\end{equation}
while for a parity odd scalar field we find
\begin{equation}
    (\Sigma_s)^-=\frac{1}{64\pi^2}\ln \left( \frac{\Lambda^2}{\mu^2} \right) \ p_5p'_5\ c_3\,.
\end{equation}
The corresponding $c_1,~c_2$ and  $c_3$ coefficients are:
\begin{equation}
    c_1=(5-\xi)g_5^2 T(r_s)\,, \qquad c_2=(1+\xi)g_5^2T(r_s)\mp\frac{\lambda_5}{2}\,, \qquad c_3=8 g_5^2 T(r_s) \pm \frac{\lambda_5}{2}\,,
\end{equation}
where $\lambda_5$ is the scalar quartic coupling $\mathcal{L}_5 \supset -\lambda_5 \left( \Phi^{\dagger} \Phi \right)^2$~\footnote{Depending on the representation of the scalar field, more quartics can be present, with a different gauge structure. Their effect must be computed case by case.}. The signs in front of the quartic contribution correspond to parity even and odd scalars  running in the purely scalar loop in Figure \ref{fig:scalar-loops}.

\begin{figure}[tbh]
\centering
\includegraphics[width=0.7\textwidth]{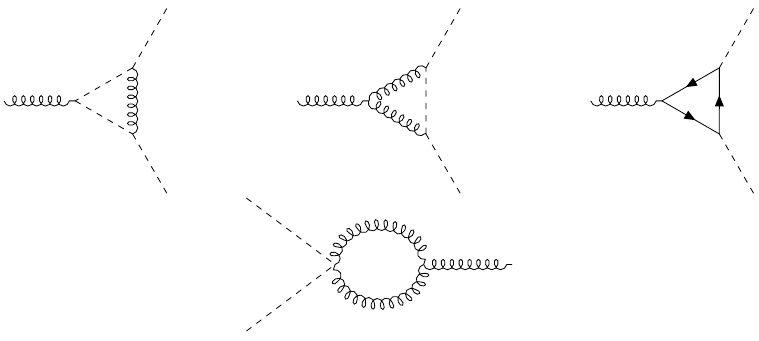}
    \caption{One loop diagrams contributing to the scalar-gauge boson vertex correction.}
    \label{fig:scalargaugeloops}
\end{figure}

There are also localized divergent contributions corresponding to the trilinear coupling of the scalars with a gauge boson. They appear only for even scalars and gauge bosons in $\mathcal{H}$. The corresponding diagrams are represented on Figure \ref{fig:scalargaugeloops}. The vertex correction has the following form:
\begin{equation}
    (\Sigma_s^{\prime})^+= \frac{g_5^2}{64 \pi^2} \ln \left( \frac{\Lambda^2}{\mu^2}  \right)\  g_5 p^{\mu} T^A c_1'\,.
\end{equation}
A similar term is also generated with two gauge bosons.

As before, a gauge-invariant set of counterterms can be obtained in the ``{\it magic}'' gauge $\xi=-3$, leading to
\begin{equation}
    c_1 = c'_1 =8g_5^2 T(r_s)\,, \qquad c_2=-2g_5^2T(r_s)\mp\frac{\lambda_5}{2}\,, \qquad c_3=8 g_5^2 T(r_s) \mp \frac{\lambda_5}{2}\,.
\end{equation}

\begin{figure}[tbh]
\centering
\includegraphics[width=0.7\textwidth]{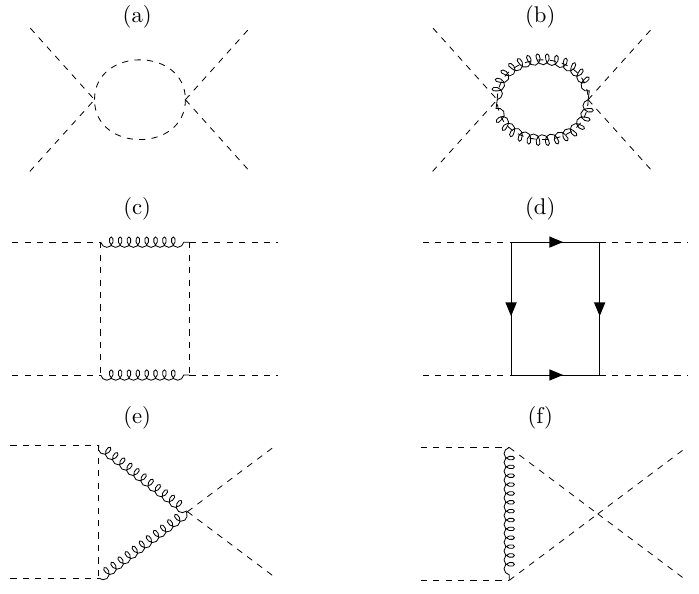}
    \caption{One loop diagrams contributing to the correction of the scalar quartic coupling $\lambda_d$.}
    \label{fig:scalar-4loops}
\end{figure}

Another correction which needs to be taken into account is the one for the scalar quartic coupling, given by the diagrams in Figure~\ref{fig:scalar-4loops}. Assuming we have an interaction of the type:
\begin{equation}
    \frac{1}{4!}\lambda_5^{ijkl}\Phi_i\Phi_j\Phi_k\Phi_l\,,
\end{equation}
and using naive dimensional analysis, we can conclude which of the one loop topologies lead to divergent contributions.
More precisely, it is all of the loops $(a)-(f)$ containing the 4D gauge bosons $A_{\mu}$, and additionally topology $(b)$ with $A_5$ propagating in the loop. Making use of the usual 4D result \cite{Cheng74,MACHACEK1985}, we find the logarithmically divergent contribution to be in the Feynman gauge $\xi=1$:
\begin{equation}
    \Pi^{\rm div}_{ijkl}=\frac{1}{16\pi^2}\ln \left( \frac{\Lambda^2}{\mu^2} \right)\left[\Lambda^2_{ijkl}-3g_5^2\Lambda^{S}_{ijkl}+4g_5^4 A_{ijkl}\right]\,,
    \label{eq:scalar_divergence_5D}
\end{equation}
where
\begin{equation}
    \begin{cases}
      \Lambda^2_{ijkl}=\lambda_5^{ijmn}\lambda_5^{mnkl}+\lambda_5^{ikmn}\lambda_5^{mnjl}+\lambda_5^{ilmn}\lambda_5^{mnjk}\,,\\
      \Lambda^S_{ijkl}=4C_2(r_s)\lambda_5^{ijkl}\,,\\
      A_{ijkl}=\{T^A,T^B\}_{ij}\{T^A,T^B\}_{kl}+\{T^A,T^B\}_{ik}\{T^A,T^B\}_{jl}+\{T^A,T^B\}_{il}\{T^A,T^B\}_{jk}\,.
    \end{cases}
\end{equation}
Notice the absence of Yukawa couplings in Eq.\eqref{eq:scalar_divergence_5D}, as summing over even and odd fermionic modes gives an overall zero contribution.
From Eq.~\eqref{eq:scalar_divergence_5D} we can see that the only change in the divergent contribution, compared to the usual 4D result, lies in the prefactor of the last term, stemming from the $g_5^4$-loops. As a result, renormalizing the scalar quartic coupling follows the same procedure as in 4D, while keeping track of this coefficient change.
The precise result, however, depends on the gauge structure of the quartics, on the number of independent couplings, and finally the result should be computed in the $\xi=-3$ gauge. Hence, the final result is model dependent.

\subsection{Summary: the localized Lagrangian $\mathcal{L}_4$} \label{sec:summary}
\paragraph{}
The one loop computation of the localized divergences in the gauge, fermion and scalar sectors performed in the previous sections allowed us to determine the full set of boundary counterterms required by renormalization. As it was mentioned above, only logarithmic divergences appear on the boundary, and only for operators that are renormalizable from a $4$D point of view. They arise only for fields associated with the unbroken subgroup $\mathcal{H}$, according to the orbifold projection. The one loop renormalized Lagrangian localized on the boundaries is given by:

\begin{eqnarray}
    \mathcal{L}_4 &=& - \frac{A}{4} F_\mathcal{H}^{\mu\nu} F_{\mathcal{H},\mu\nu} + B_1\ i\bar{\psi} \cancel{\mathcal{D}} \frac{1\pm\gamma_5}{2} \psi + B_2 \left[ (\partial_5 \bar{\psi}) \frac{1\pm\gamma_5}{2} \psi - \bar{\psi} \frac{1\mp\gamma_5}{2} (\partial_5 \psi)\right] + \nonumber\\
    && + C_1\ (\mathcal{D}_\mu \phi_{+})^\dagger (\mathcal{D}^\mu \phi_{+}) + C_2 \left[(\partial_5^2 \phi_{+}^\dagger) \phi_{+} + \phi^\dagger_{+} (\partial_5^2 \phi_{+}) \right] + C_3\ (\partial_5 \phi_{-})^\dagger (\partial_5 \phi_{-}) + \nonumber \\
    && - m_\text{loc}^2 (\phi_{+}^\dagger \phi_{+}) - D\ (\phi_{+}^\dagger \phi_{+})^2\,,
\end{eqnarray}
where $\phi_{\pm}$ indicate parity even and odd scalar components, respectively. The covariant derivative $\mathcal{D}$ only contains the unbroken generators in $\mathcal{H}$. All the coefficients $A$, $B_i$, $C_i$ and $D$ are fixed by the localized divergences computed above. Matching the one-loop computations in the ``{\it magic}'' gauge $\xi=-3$:
\begin{subequations}\label{eq:loccounterterms}
\begin{eqnarray}
    A &=& \frac{g_5^2}{64 \pi^2} \ln \left( \frac{\Lambda^2}{\mu^2} \right) \left[ \frac{23}{3}(2 C(\mathcal{H}) - C(\mathcal{G}))-\frac{1}{3} \left( \sum_{\rm even}-\sum_{\rm odd}  \right) T(r_s) \right] \,, \\
    B_1 &=& \frac{1}{64 \pi^2} \ln \left( \frac{\Lambda^2}{\mu^2} \right) \left[ 7g_5^2 C(r)- \left( \sum_{\rm even}-\sum_{\rm odd} \right)y_5^2  \right] \,, \\
    B_2 &=&  \frac{1}{64 \pi^2} \ln \left( \frac{\Lambda^2}{\mu^2} \right) \left[ g_5^2 C(r)- \left( \sum_{\rm even}-\sum_{\rm odd} \right)y_5^2  \right] \,, \\
    C_1 &=&  \frac{1}{64 \pi^2} \ln \left( \frac{\Lambda^2}{\mu^2} \right) \left[ 8g_5^2 T(r)  \right]\,, \\
    C_2 &=& \frac{1}{64 \pi^2} \ln \left( \frac{\Lambda^2}{\mu^2} \right) \left[ -2g_5^2 T(r) \mp \frac{\lambda_5}{2}  \right] \,, \\
    C_3 &=& \frac{1}{64 \pi^2} \ln \left( \frac{\Lambda^2}{\mu^2} \right) \left[ 8g_5^2 T(r) \mp \frac{\lambda_5}{2}  \right] \,, 
\end{eqnarray}
\end{subequations}
while the scalar quartic couplings $D$ depend on the specificity of the model and on the scalar gauge representations.
Note that the couplings appearing in the above formulae are still the dimension-full 5D ones. As a result, all the coefficients have dimension $m^{-1}$, which in turn guarantees that all operators in $\mathcal{L}_4$ have dimension $m^4$, as it should be from its definition in Eq.~\eqref{eq:S5}.

\section{RGE evolution and the existence of fixed points} \label{sec:bulkRGEs}
\begin{figure}[tbh]
\centering
\includegraphics[width=0.5\textwidth]{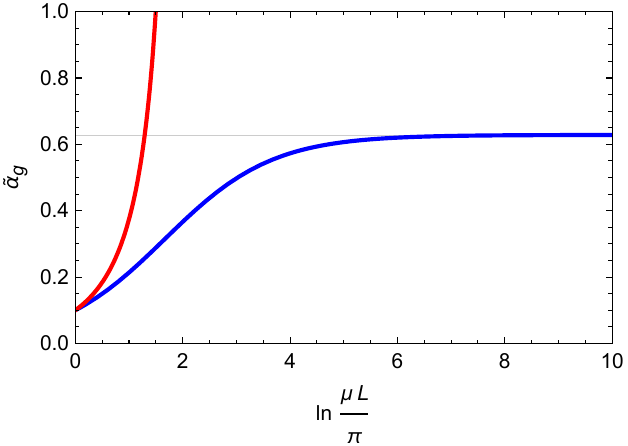}
\caption{Example of running of the effective 't Hooft gauge coupling $\tilde{\alpha}_g$ for a generic 5D theory with power-law running (red with $b_5 = -10$) and with a fixed point (blue with $b_5=10$).} \label{fig:runningplot}
\end{figure}

\paragraph{}
Even though the 5D bulk theory described by a Lagrangian of the form of Eq.~\eqref{eq:LagrGen} may be formally renormalizable, its validity could be limited by the renormalization running of the couplings. In fact, as they carry mass dimension, they run as a power law, hence they may diverge in the UV, at a scale not too far from the scale of the extra dimension. This is shown, schematically, by the red curve in Fig.~\ref{fig:runningplot}. As already stressed, this problem is not present for logarithmically running couplings.

This critical issue can be avoided if the RGEs admit a UV fixed point for all the power-law running couplings in the bulk \cite{Gies_2003,Morris_2005}. An illustration is provided by the blue curve in Fig.~\ref{fig:runningplot}. Here we recap the conditions leading to UV fixed points for gauge, Yukawa and quartic scalar couplings in 5D. This allows to establish the necessary constraints on the field content of potentially renormalizable 5D models.

\subsection{Gauge couplings}
\paragraph{}
As we have seen, the 5D gauge coupling $g_5$ carries mass dimension $[g_5] = m^{-1/2}$. When the theory is compactified, it is matched to a reduced 4D gauge couplings $g = g_5/\sqrt{L}$. 
The loop factor, however, depends linearly on the energy: this can be seen as due to the mass dimension of the 5D coupling, or alternatively to the sum over the KK states below the energy $\mu$ (whose counting is given by $\mu L/\pi$, where $\pi/L=m_{\rm KK}$ is the KK mass scale). In either case, one can define a dimension-less effective 't Hooft coupling, which controls the loop corrections:
\begin{equation}
\tilde{\alpha}_g=\frac{\alpha_g\mu}{\pi}\,, \quad \text{where}\;\; \alpha_g = \frac{g_5^2}{4\pi}\,.
\end{equation}
Then, the one-loop RGE in terms of $\tilde{\alpha}_g$ at energy $\mu \gtrsim \pi/L$ reads \cite{Cacciapaglia21}:
\begin{equation}
2\pi\frac{\dd \tilde{\alpha}_g}{\dd \ln(\mu)} = 2\pi \tilde{\alpha}_g -b_5 \tilde{\alpha}_g^2\,,
\label{eq:gaugeRGE}
\end{equation}
where the first term on the right-hand-side stems precisely from the $\mu$-dependence inside $\tilde{\alpha}_g$.
The 5D one-loop beta function $b_5$ is given by:
\begin{equation}
    b_5=\frac{21}{6} C(G)-\frac{4}{3}\sum_{f}T(R_f)-\frac{1}{3}\sum_s T(R_s)\,,
\end{equation}
where the subscripts $s$ and $f$ stand for complex scalar and fermion representations in the bulk. From Eq.~\eqref{eq:gaugeRGE}, a condition for the existence of the UV fixed point can be derived. This condition reads $b_5>0$,  and the corresponding fixed point takes the following value:
\begin{equation}
    \Tilde{\alpha}^*_{g}=\frac{2\pi}{b_5}\,.
    \label{eq:gaugeFP}
\end{equation}
In Fig.~\ref{fig:runningplot} we plot a sample running of $\tilde{\alpha}_g$ for $b_5 = \pm 10$: the red curve illustrate the divergent running of theories that lose control above the KK scale (here, around ten KK states can be described as weakly coupled); the blue curve illustrate a theory with UV fixed point, where an arbitrary number of KK modes can be included perturbatively.

Since Eq.~\eqref{eq:gaugeFP} is derived perturbatively, a necessary requirement is that $\Tilde{\alpha}^*_{g}$ stays perturbative as well. 
However, one needs to take into account the extra-dimensional loop definition, such that the effective loop factor $\xi_g(d)$ is proportional to the $d$-dimensional solid angle:
\begin{equation}
    \xi_g(d)= \frac{\Omega(d)}{(2 \pi)^d} 4 \pi \Tilde{\alpha}_g\,,
\end{equation}
with $\Omega(d)= 2 \pi^{5/2} / \Gamma(d/2)$ the $d$-dimensional solid angle. In 5D we get $\xi_g(5) = \Tilde{\alpha}_g / 3 \pi^2$, so that the fixed point remains perturbative as long as $b_5 > 2/3 \pi$.

\subsection{Yukawa couplings}
\paragraph{}
The RGEs of the Yukawa couplings $y_5$ can be treated in a similar way as the gauge case. We can also define an effective 't Hooft coupling $\Tilde{\alpha}_y=\alpha_y \mu/\pi$, where $\alpha_y = y_5^2/4\pi$. The corresponding RGE reads \cite{Cacciapaglia:2023kyz}:
\begin{equation}
2\pi\frac{\dd \tilde{\alpha}_y}{\dd \ln(\mu)} = 2\pi \tilde{\alpha}_y +c_y \tilde{\alpha}_y^2-d_y\Tilde{\alpha}_g\Tilde{\alpha}_y\,.
\end{equation}
The constants $c_y$ and $d_y$ are model dependent and should be computed case-by-case at one loop.
A fixed point can be found by searching for the zeros of the beta function, which are given by:
\begin{equation}
    \Tilde{\alpha}_y^*=\frac{d_y\Tilde{\alpha}_g^*-2\pi}{c_y}\,,
    \label{eq:yukawaFP}
\end{equation}
where $\Tilde{\alpha}_g^*$ is the fixed point value of the gauge coupling in Eq.~\eqref{eq:gaugeFP}. Similarly to the gauge coupling case, the extra-dimensional loop factor needs to be taken into account to assess the perturbativity of the Yukawa coupling.
We deduce from Eq.~\eqref{eq:yukawaFP} that the Yukawa coupling has a repulsive fixed point for $c_y>0$ (as usually found) and: 
\begin{equation}
d_y\Tilde{\alpha}_g^*>2\pi\,.
\label{eq:yukawaFPcondition}
\end{equation}
Using the definition of the gauge fixed point from Eq.~\eqref{eq:gaugeFP}, we can rewrite this condition:
\begin{equation}
    d_y>b_5> \frac{2}{3\pi}\,,
\end{equation}
which checks for both a Yukawa fixed point and a gauge perturbative one.
When the theory features multiple Yukawa couplings in the bulk, a set of linked RGEs must be solved \cite{Cacciapaglia:2023kyz}:
\begin{equation}
  2\pi\frac{\dd \tilde{\alpha}_y}{\dd \ln(\mu)} = \left(2\pi  +\sum_{y'}c_{yy'} \tilde{\alpha}_{y'}-d_{y}\Tilde{\alpha}_g\right)\Tilde{\alpha}_y\,.  
\end{equation}
The zeros of the beta-function depend now on the inverse matrix $c^{-1}_{yy'}$:
\begin{equation}
    \Tilde{\alpha}_y^*=\sum_{y'}c^{-1}_{yy'}\left({d_{y'}\Tilde{\alpha}_g^*-2\pi}\right)\,.
    \label{eq:yukawaFPmatrix}
\end{equation}
As such, a complete fixed point exists if all solutions in Eq.~\eqref{eq:yukawaFPmatrix} are positive.

\subsection{Scalar couplings}
\paragraph{}
In the presence of bulk scalar fields, the 5D theory must also contain scalar quartic interactions, schematically given in terms of a coupling $\lambda_5$ of dimension $m^{-1}$. Similarly to the previous cases, we can define a dimensionless quantity $\Tilde{\alpha}_{\lambda}=\frac{\lambda_5}{4\pi^2}\mu$, such that the RGE reads:
\begin{equation}
    4\pi\frac{\dd \Tilde{\alpha}_{\lambda}}{\dd \ln(\mu)}=4\pi\Tilde{\alpha}_{\lambda}+a_{\lambda}\Tilde{\alpha}^2_{\lambda}-b_{\lambda}\Tilde{\alpha}^2_{y}+d_{\lambda}\Tilde{\alpha}^2_{g}-c_{\lambda}\Tilde{\alpha}_g\Tilde{\alpha}_{\lambda}+e_{\lambda}\Tilde{\alpha}_{\lambda}\Tilde{\alpha}_{y}\,.
    \label{eq:scalarRGE}
\end{equation}
Once again, the model dependent coefficients $\{a_{\lambda},...,e_{\lambda}\}$ arise from the explicit one loop computations.
The fixed points, given by the zeros of the beta-function, read:
\begin{equation}
  \Tilde{\alpha}_{\lambda}^*=\frac{-(4\pi-c_{\lambda}\Tilde{\alpha}^*_{g}+e_{\lambda}\Tilde{\alpha}^*_{y})\pm\sqrt{(4\pi-c_{\lambda}\Tilde{\alpha}^*_{g}+e_{\lambda}\Tilde{\alpha}^*_{y})^2-4a_{\lambda}(-b_{\lambda}\Tilde{\alpha}^{*2}_{y}+d_{\lambda}\Tilde{\alpha}^{*2}_{g})}}{2a_{\lambda}} \,.
  \label{eq:scalarFP}
\end{equation}
We notice the existence of two solutions due to the quadratic nature of the beta-function in the effective coupling.
The existence of the fixed points is guaranteed as long as the solutions are real, i.e. the argument of the square root is positive, leading to the constraint
\begin{equation}
    (4\pi - c_\lambda \tilde{\alpha}_g^\ast + e_\lambda \tilde{\alpha}_y^\ast)^2 > 4 a_\lambda (d_\lambda \tilde{\alpha}_g^{\ast 2} - b_\lambda \tilde{\alpha}_y^{\ast 2})\,.
\end{equation}
Inspecting Eq.~\eqref{eq:scalarRGE} shows that as long as there exists a bulk scalar field $\phi$ in the theory, a bulk scalar coupling $\lambda_5$ will be radiatively generated. This is a consequence of the presence of $b_{\lambda}$ and $d_{\lambda}$ terms, which are strictly Yukawa and gauge, respectively. As a result, there are no purely localized scalar quartic couplings.

The structure of the fixed points is more complicated in the presence of multiple Yukawas and multiple quartics, and it should be analyzed case-by-case. Note also that the stability of the potential at the fixed point may impose additional constraints on the model.

\section{Fixed points and RGEs for localized interactions}\label{sec:RGEloc}
\paragraph{}
As we have seen in Section~\ref{sec:locdiv}, localized divergences require the presence of localized interactions in $\mathcal{L}_4$. These terms arise as counterterms of logarithmic divergences. They can also be seen as the effect of logarithmic running from bulk couplings, hence their running does not jeopardize the validity of the theory. 
Nevertheless, theories may include purely localized terms made of bulk fields. Such couplings cannot be  gauge nor scalar quartics, since they always originate as counterterms from bulk interactions. Henceforth, only new Yukawa couplings shall be considered, as they do not arise as localized counterterms.

To this end, we assume the presence of a localized interaction of the form:
\begin{equation}
    \delta \mathcal{L}_{4}= y_{\rm loc} \psi\psi\phi\,,
\end{equation}
made of parity-even components of bulk fields, and invariant under $\mathcal{H}$. Hence, $y_{\rm loc}$ has the same mass dimension as a bulk Yukawa, $[y_{\rm loc}]=m^{-1/2}$. One can therefore construct a 4D effective coupling $y_{4,\text{loc}} = y_{\rm loc}/\sqrt{L}$ and $\alpha_{y4}=y_{4,\text{loc}}^2/(4\pi)$. The RGE can be extracted from the usual one loop diagrams,
with at least one coupling replaced by $y_{\rm loc}$. The degree of divergence can be easily estimated by working in KK expansion, recalling that bulk couplings conserve KK momentum, while localized ones do not. Considering the contribution of the localized Yukawa alone, the RGE for the 4D coupling $\alpha_{y4}$ read:
\begin{equation}
    2\pi \frac{\dd \alpha_{y4}}{\dd \ln(\mu)} = c_{y}^{\rm loc} \alpha_{y4}^2\ \left(\frac{\mu L}{\pi}\right)^3\,,
\end{equation}
where we accounted for three sums over KK states with masses below $\mu$. The $\mu$-dependence can be reabsorbed by defining an effective 't Hooft coupling:
\begin{equation}
    \tilde{\alpha}_{y_{\rm loc}}=\frac{y^2_{\rm loc}}{4\pi L}\left(\frac{\mu L}{\pi}\right)^3\,.
\end{equation}
Hence, the RGE reduces to
\begin{equation}
    2 \pi \frac{\dd \tilde{\alpha}_{y_{\rm loc}}}{\dd \ln(\mu)} = 6\pi \tilde{\alpha}_{y_{\rm loc}} + c_y^{\rm loc} \tilde{\alpha}_{y_{\rm loc}}^2\,.
\end{equation}

In a similar manner, one can include the loops containing bulk gauge and Yukawa couplings. For multiple localized Yukawas, the complete RGEs read:
\begin{equation}
    2\pi\frac{\dd \Tilde{\alpha}_{y_{\rm loc}}}{\dd \ln(\mu)}=6\pi\Tilde{\alpha}_{y_{\rm loc}}+ \Tilde{\alpha}_{y_{\rm loc}} \sum_{y'}c_{yy'}^{\rm loc} \Tilde{\alpha}_{y'_{\rm loc}}+\Tilde{\alpha}_{y_{\rm loc}}\left[-d_y\Tilde{\alpha}_g+f_y\Tilde{\alpha}_{y}\right]\left(\frac{\mu L}{\pi}\right)\,.
    \label{eq:locYukRGE}
\end{equation}
In the above notation $\Tilde{\alpha}_{y}$ and $\Tilde{\alpha}_{g}$ are the bulk Yukawa and gauge effective couplings. The $\mu$-dependence is now explicit, through the last term in Eq.~\eqref{eq:locYukRGE}. For large $\mu$, it is this term that dominates, hence depending on its sign, $\tilde{\alpha}_{y_{\rm loc}}$ will either diverge or run to zero. Hence, for 
\begin{equation}
    d_y \tilde{\alpha}_g^\ast - f_y \tilde{\alpha}_y^\ast > 0
\end{equation}
the localized Yukawa will run to a Gaussian fixed point in the UV at $\Tilde{\alpha}_{y_{\rm loc}}^\ast = 0$.

The necessary condition for an interactive fixed point is given by:
\begin{equation}
-d_y\Tilde{\alpha}^*_g+f_y\Tilde{\alpha}^*_{y}=0\,.    
\end{equation}
If the condition is satisfied at the fixed points of the gauge and bulk Yukawa, then the localized Yukawas will flow to interactive fixed points at:
\begin{equation}
  \Tilde{\alpha}_{y_\text{loc}}^*=-6\pi\sum_{y'} (c^{\rm loc}_{yy'})^{-1}\,,
  \label{eq:locYukFP}
\end{equation}
which exist if all solutions are positive.

\section{Summary and outlook} \label{sec:concl}
\paragraph{}
In this work we carried out a systematic analysis of one loop renormalization of five-dimensional gauge-Yukawa theories compactified on the $\mathcal{S}_1/ \mathbb{Z}_2$ orbifold, with particular emphasis on the interplay between bulk and localized divergences. Our study demonstrates that, despite the non-renormalizable character expected from naive power counting, 5D gauge-Yukawa theories share the same divergent operator structure as their 4D counterparts at one loop, both in the bulk and on the boundaries. All bulk divergences can be reabsorbed into a minimal 5D Lagrangian made of operators that would be renormalizable in 4D. Localized logarithmic divergences require 4D-renormalizable operators on the boundaries, showing that the full counterterm structure remains finite and under control. Explicit computations for gauge bosons, fermions and scalars reveal that localized divergences arise only for fields associated with the subgroup $\mathcal{H}$ preserved by the orbifolding on $\mathcal{S}_1/ \mathbb{Z}_2$, and that in the ``{\it magic}'' $\xi=-3$ gauge they can be consistently organized into gauge invariant combinations. The complete set of of localized counterterms is summarized in the effective boundary Lagrangian $\mathcal{L}_4$. 

\paragraph{}
Building on this framework, we analyzed the RGE evolution of bulk and localized couplings affected by power-law divergences. Using the effective 't Hooft couplings appropriate to $5$D theories, we established the conditions for the existence of UV fixed points for gauge, Yukawas and scalar quartic interactions. These fixed points arise naturally due to the characteristic power-law running induced by the Kaluza-Klein tower. 
These results are of particular interest for 5D Grand unification theories, where gauge couplings flow towards the same UV fixed point. Conditions for the existence of the fixed points can be derived, depending on the field content of the model under discussion. If such fixed points are present, and the high-energy behavior is under control, then the question of renormalizability of 5D models becomes crucial. In fact, for the asymptotic unification to be theoretically under control, it suffices that the power-law divergences are absorbed in a finite number of counterterms. Logarithmic divergences are less threatening as they may spoil the theory at scales much higher than the compactification one, in a regime where the bulk couplings are close to the UV fixed points and new physics related to quantum gravity is needed.

\paragraph{}
Our results offer a coherent picture to define 5D gauge-Yukawa theories valid up to high scales. Our results also define a concrete toolbox for the realizations of phenomenologically interesting models, such as asymptotic Grand unified theories (aGUTs) or any extra-dimensional extension of the Standard Model. Extending the present analysis to higher loops is non-trivial as it requires a detailed analysis of the emergence of new power-law divergences. Hence, this remains a compelling direction for future work and is crucial for establishing the full renormalizability properties of 5D gauge-Yukawa theories.

\section*{Acknowledgements}

We thank Aldo Deandrea, Florian Nortier, Alan Cornell, Shao Hua-Sheng and Wang Zhi-Wei for useful discussion and suggestions, relevant for this work.

\bibliographystyle{utphys}
\bibliography{bib}

\end{document}